\documentclass[aps,prd,twocolumn,superscriptaddress,floatfix]{revtex4-2}

\usepackage{graphicx}
\usepackage{tikz}
\usepackage{xcolor}
\usepackage{amsmath,amssymb,amsfonts}
\usepackage{comment}
\usepackage{float}
\usepackage{placeins}
\usepackage[colorlinks=true, linkcolor=blue, urlcolor=blue, citecolor=blue]{hyperref}
\usepackage{breakurl}
\usepackage{cleveref}
\usepackage{siunitx}
\usepackage[latin1]{inputenc}
\usepackage{bm}
\usepackage{physics}
\usepackage{empheq}
\usepackage{ulem}
\usepackage{orcidlink}

\newcommand{\gag}{g_{a \gamma}}
\renewcommand{\Re}{\operatorname{Re}}
\renewcommand{\Im}{\operatorname{Im}}

\begin{document}

\tikz[remember picture, overlay] \node[anchor=north east, xshift=-1cm, yshift=-1cm] at (current page.north east) {\small Preprint: KEK-TH-2720};

\title{Adjusting optical cavity birefringence with wavelength tunable laser for axion searches}

\author{Hinata~Takidera\orcidlink{0009-0007-2482-1320}}
\affiliation{Department of Physics, The University of Tokyo, Bunkyo, Tokyo 113-0033, Japan}
\author{Hiroki~Fujimoto\orcidlink{0000-0002-0070-0678}}
\affiliation{Department of Physics, The University of Tokyo, Bunkyo, Tokyo 113-0033, Japan}
\author{Yuka~Oshima\orcidlink{0000-0002-1868-2842}}
\affiliation{Department of Physics, The University of Tokyo, Bunkyo, Tokyo 113-0033, Japan}
\author{Satoru~Takano\orcidlink{0000-0002-1266-4555}}
\affiliation{Department of Physics, The University of Tokyo, Bunkyo, Tokyo 113-0033, Japan}
\affiliation{Max-Planck-Institut f\"{u}r Gravitationsphysik (Albert-Einstein-Institut), D-30167 Hannover, Germany}
\author{Kentaro~Komori\orcidlink{0000-0002-4092-9602}}
\affiliation{Department of Physics, The University of Tokyo, Bunkyo, Tokyo 113-0033, Japan}
\affiliation{Research Center for the Early Universe (RESCEU), Graduate School of Science, The University of Tokyo, Bunkyo, Tokyo 113-0033, Japan}
\author{Tomohiro~Fujita\orcidlink{0000-0003-4722-7432}}
\affiliation{Department of Physics, Ochanomizu University, Bunkyo, Tokyo 112-8610, Japan}
\affiliation{Research Center for the Early Universe (RESCEU), Graduate School of Science, The University of Tokyo, Bunkyo, Tokyo 113-0033, Japan}
\affiliation{Kavli Institute for the Physics and Mathematics of the Universe (Kavli IPMU), WPI, UTIAS, The University of Tokyo, Kashiwa, Chiba 277-8568, Japan}
\author{Ippei~Obata\orcidlink{0000-0001-9737-5631}}
\affiliation{Kavli Institute for the Physics and Mathematics of the Universe (Kavli IPMU), WPI, UTIAS, The University of Tokyo, Kashiwa, Chiba 277-8568, Japan}
\affiliation{Theory Center, Institute of Particle and Nuclear Studies (IPNS), High Energy Accelerator Research Organization (KEK), 1-1 Oho, Tsukuba, Ibaraki 305-0801, Japan}
\author{Masaki~Ando\orcidlink{0000-0002-8865-9998}}
\affiliation{Department of Physics, The University of Tokyo, Bunkyo, Tokyo 113-0033, Japan}
\affiliation{Research Center for the Early Universe (RESCEU), Graduate School of Science, The University of Tokyo, Bunkyo, Tokyo 113-0033, Japan}
\author{Yuta~Michimura\orcidlink{0000-0002-2218-4002}}
\affiliation{Research Center for the Early Universe (RESCEU), Graduate School of Science, The University of Tokyo, Bunkyo, Tokyo 113-0033, Japan}
\affiliation{Kavli Institute for the Physics and Mathematics of the Universe (Kavli IPMU), WPI, UTIAS, The University of Tokyo, Kashiwa, Chiba 277-8568, Japan}

\begin{abstract}
    Axions have attracted attention as promising candidates for dark matter (DM).
    Although axions have been intensively searched for, they have not been observed yet.
    Recently, novel experiments to search for axion DM have been proposed that use optical cavities to amplify polarization rotation of laser light induced by the axion-photon interaction.
    One such experiment employs a ring cavity composed of four mirrors.
    However, its sensitivity to the axion-photon coupling $\gag$ in the low axion mass region is limited due to a reflection phase difference between s- and p-polarizations.
    In this paper, we propose a new method to improve the sensitivity using zero phase shift mirrors and a wavelength tunable laser.
    Moreover, the laser makes it easier to scan the high axion mass region by tuning the reflection phase difference between s- and p-polarizations.
    We experimentally confirmed that the phase difference generated upon reflection on a zero phase shift mirror satisfied the requirement of $8.6 \times 10^{-3}~\si{deg}$, which corresponds to the half width at half maximum (HWHM) of the cavity for p-polarization with the mirror fixed on a folded cavity and a wavelength tunable laser.
\end{abstract}

\maketitle

\section{Introduction}
Axions are one of the leading candidates for dark matter (DM)~\cite{PRESKILL_1983_127, ABBOTT_1983_133, DINE_1983_137}.
They were originally proposed to solve the strong \textit{CP} problem in quantum chromodynamics (QCD)~\cite{PhysRevLett.38.1440}, and are generally called the QCD axion.
String theory and supergravity suggest the existence of axionlike particles (ALPs) with a wide range of mass~\cite{Svrcek_2006, PhysRevD.81.123530}.
Although ALPs do not solve the strong \textit{CP} problem, they are nonetheless well-motivated candidates for DM.
In this paper, we collectively call the QCD axion and ALPs as ``axions."

Axions are ultralight particles that weakly interact with photons, electrons, protons, and neutrons.
Many experiments have searched for axions utilizing the axion-photon interaction with magnetic field.
For example, axion haloscopes~\cite{PhysRevLett.51.1415, PhysRevD.64.092003} (ADMX~\cite{PhysRevLett.104.041301, PhysRevLett.120.151301, PhysRevLett.124.101303, PhysRevLett.127.261803}, ADMX SLIC~\cite{PhysRevLett.124.241101}), axion helioscopes~\cite{PhysRevLett.69.2333} (Sumico~\cite{MORIYAMA_1998_147, INOUE_2008_93}, CAST~\cite{CAST_2017, PhysRevLett.133.221005}), light shining through a wall experiments~\cite{PhysRevLett.59.759, PhysRevD.37.2001} (ALPS~\cite{EHRET_2009_83, EHRET_2010_149}, OSQAR~\cite{PhysRevD.92.092002}), PVLAS experiment~\cite{Valle_2016}, measurements of magnetic field oscillations under static magnetic field (ABRACADABRA~\cite{PhysRevLett.117.141801, PhysRevLett.127.081801}, SHAFT~\cite{Gramolin_2021}), astronomical observations (SN1987A~\cite{BROCKWAY_1996_439, Payez_2015}, M87~\cite{Marsh_2017}, NGC1275~\cite{Reynolds_2020}, H1821+643~\cite{H1821643_2021}, pulsars~\cite{PhysRevLett.131.111004}, axion star explosions~\cite{PhysRevD.109.043018}, CMB~\cite{PhysRevLett.134.081001}, MWD polarization~\cite{benabou2025searchaxionsmagneticwhite}, M82~\cite{PhysRevLett.134.171003}), and Earth's magnetic fields (SNIPE~\cite{PhysRevD.108.096026}, SuperMAG~\cite{PhysRevD.105.095007, PhysRevD.110.115036}, Eskdalemuir~\cite{nishizawa2025axiondarkmattersearch}) have set upper limits.
Although axions have been intensively searched for, they have not been discovered yet.

Recently, novel experiments to search for axion DM have been proposed that utilize the axion-photon interaction without magnetic field.
These experiments aim to detect a phase velocity difference between left- and right-handed circularly polarized light using an optical cavity~\cite{PhysRevD.98.035021}.
When the basis of circularly polarized light is converted to that of linearly polarized light, it can be regarded as a polarization rotation.
Gravitational wave detectors can be used to search for axion DM by detecting the polarization rotation~\cite{PhysRevLett.123.111301}.
In this method, using the arm cavity reflection ports of gravitational wave detectors degrades the sensitivity to the axion-photon coupling in the low axion mass region due to the flip of the polarization plane upon reflection on the mirror.
On the other hand, using the transmission ports improves the sensitivity in the same mass region since photons travel in the cavity an odd number of times~\cite{PhysRevD.104.062008}.
LIDA~\cite{PhysRevLett.132.191002} and ADBC experiment~\cite{PhysRevD.100.023548, PhysRevLett.133.111003} are laser-interferometric detectors with ring cavities searching for axion DM.
The ring cavity prevents cancellation of the polarization rotation angle caused by the flip of the polarization plane upon reflection on the mirror.
LIDA~\cite{PhysRevLett.132.191002} and ADBC experiment~\cite{PhysRevLett.133.111003} have searched for axion DM near \SI{2}{neV} and \SI{50}{neV}, respectively.

We proposed the Dark matter Axion search with riNg Cavity Experiment (DANCE)~\cite{PhysRevLett.121.161301}.
We inject linearly polarized light into a bow-tie ring cavity and amplify the polarization rotation angle due to the axion-photon interaction.
One of the main challenges faced by experiments using a ring cavity is that the resonant frequency difference between s- and p-polarizations degrades the sensitivity in the low axion mass region~\cite{PhysRevLett.133.111003, PhysRevLett.132.191002, PhysRevD.108.072005}, in contrast to experiments using a linear cavity such as \cite{PhysRevD.98.035021}, \cite{PhysRevLett.123.111301}, and \cite{PhysRevD.104.062008}.
The parity inversion caused by mirror reflection leads to a redefinition of the polarization axis.
Independent of this effect, the reflection phase difference between s- and p-polarizations induced upon reflection on the mirror is ideally \SI{0}{deg}, under the assumption that the reflectivity is ideal.
However, in a ring cavity configuration where the laser light is incident on the mirrors at an oblique angle, this phase difference shifts from \SI{0}{deg} due to the dielectric multilayer coating of the mirror.
The phase difference generates the resonant frequency difference between s- and p-polarizations.
The detail is described in Sec.~\ref{sec:principle}.

Here, we define simultaneous resonance as the case where the resonant frequency difference between s- and p-polarizations is smaller than the half width at half maximum (HWHM) of the cavity for axion signal.
Conversely, we define nonsimultaneous resonance as the case where the frequency difference is larger than the HWHM.
In order to conduct a sensitive axion DM search in the low axion mass region, achieving simultaneous resonance is essential.
According to the theoretical calculation of the mirror coating, the reflection phase difference between s- and p-polarizations generated by mirror reflection depends on both the angle of incidence on the mirror and the laser wavelength.
One approach to achieving simultaneous resonance is tuning the phase difference by adjusting the angle of incidence through mirror rotation~\cite{PhysRevD.100.023548}.
However, the results have some gaps in the searched axion mass range due to manually rotating the mirrors~\cite{PhysRevLett.133.111003}.
To search over a wide axion mass range, the mirrors need to be linearly actuated with an accuracy on the order of \SI{0.1}{mm}.
The other approach is using an auxiliary cavity~\cite{PhysRevD.101.095034}.
In this method, the p-polarization is resonant with a ring cavity, and the s-polarization is resonant with a ring cavity and an auxiliary cavity.
The auxiliary cavity compensates for the resonant frequency difference between s- and p-polarizations.
We presented a method to achieve simultaneous resonance by incorporating an auxiliary cavity to a bow-tie ring cavity~\cite{Fujimoto_2021}.
However, the optical loss generated on the polarizing beam splitter (PBS) between the bow-tie ring cavity and the auxiliary cavity degraded the sensitivity.

In this paper, we propose a new method for a broadband and sensitive axion DM search.
The approach is to use zero phase shift mirrors and a wavelength tunable laser.
We can tune the reflection phase difference between s- and p-polarizations by varying the laser wavelength without the auxiliary cavity.
In this work, we designed a folded cavity to demonstrate a proof of principle for simultaneous resonance and to evaluate the reflection phase difference between s- and p-polarizations per mirror with the zero phase shift mirror and the wavelength tunable laser.
In Sec.~\ref{sec:principle}, we describe the principle of axion DM search and the reflection phase difference between s- and p-polarizations.
Section~\ref{sec:experiment} outlines the concept and setup of our experiment.
In Sec.~\ref{sec:results} and \ref{sec:discussion}, we present and discuss the results of this work.
Finally, Sec.~\ref{sec:conclusion} provides the conclusion.
In this paper, we set the natural unit system, $\hbar=c=1$, unless otherwise noted.

\section{Principle}
\label{sec:principle}
Assuming that DM is composed of ALPs, due to its large number density, the axion field $a(t)$ behaves like a classical wave with axion mass $m_a$, and is expressed by
\begin{equation}
    a(t) = a_0 \cos{\qty(m_a t + \delta_{\tau}(t))},
\end{equation}
where $a_0$ is the axion field amplitude and $\delta_{\tau}(t)$ is the phase factor.
The axion field oscillate with a frequency $f_a$, which is related to the axion mass $m_a$ by $f_a=m_a/(2\pi)=\SI{242}{Hz}\,(m_a/{10^{-12}}~\si{\eV})$.
$\delta_{\tau}(t)$ stays constant during the coherent time $\tau=\lambda_a/v_a=2\pi/m_a v_a^2$, where $\lambda_a=2\pi/m_a v_a$ is the de Broglie wavelength of axion DM, $v_a\sim10^{-3}$ is the galactic virial velocity of axion DM~\cite{BERTONE_2005_279}.
The phase velocity of left- and right-handed circularly polarized light is given by
\begin{gather}
    c_\mathrm{L/R}(t) = 1 \pm \delta c_0 \sin{\qty(m_a t + \delta_{\tau}(t))}, \\
    \delta c_0 \equiv \frac{\gag \sqrt{2 \rho_a}}{2k_0},
\end{gather}
where $\gag$ is the axion-photon coupling, $\rho_a \equiv m_a^2 a_0^2/2 \sim \SI{0.4}{GeV~cm^{-3}}$ is the energy density of axion DM in the solar system~\cite{Weber_2010, RiccardoCatena_2010}, and $k_0$ is the wave number of laser light.

\begin{figure}[H]
    \centering
    \includegraphics[width=1.0\linewidth]{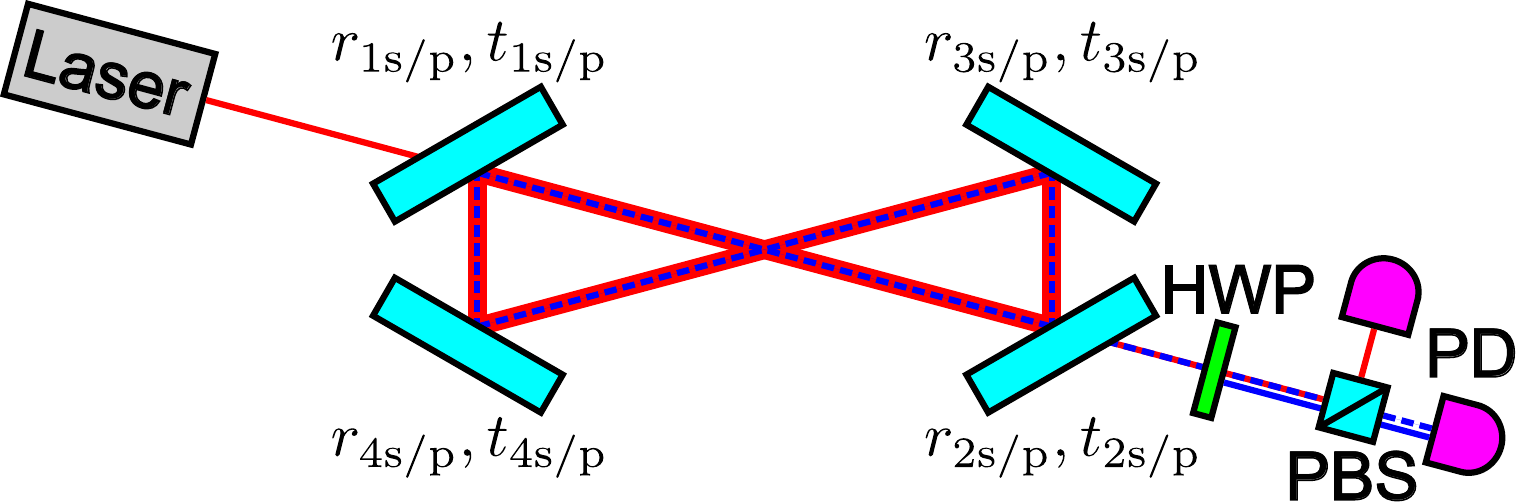}
    \caption{
    Configuration of DANCE.
    The red and blue solid lines indicate s- and p-polarizations, respectively.
    The blue dashed line represents the p-polarization generated via the axion-photon interaction.
    $r_{j\mathrm{s/p}}$ and $t_{j\mathrm{s/p}}$ are the amplitude reflectivity and transmissivity, respectively, for s- and p-polarizations of the $j$th mirror $(j=1,2,3,4)$.
    HWP: half-wave plate,
    PBS: polarizing beam splitter,
    PD: photodetector.
    }
    \label{fig:DANCE}
\end{figure}

Recently, we developed DANCE~\cite{PhysRevD.108.072005}.
If we transform the basis of circularly polarized light to that of linearly polarized light, it is interpreted as a polarization rotation.
When the polarization is adjusted to s-polarization, the polarization rotation induces p-polarization sidebands.
In the following, we consider linearly polarized light.
Figure~\ref{fig:DANCE} shows the configuration of DANCE, which consists of four mirrors.
We number the mirrors in the order of the light propagation direction as $j=1,2,3,4$.
Here, we define the amplitude reflectivity and transmissivity of the $j$th mirror for s- and p-polarizations as $r_{j\mathrm{s}}, t_{j\mathrm{s}}$, and $r_{j\mathrm{p}}, t_{j\mathrm{p}}$, respectively.
When we inject s-polarization into the cavity as shown in Fig.~\ref{fig:DANCE}, the electric field of the transmitted light $\bm{E}_\mathrm{t}(t)$ is given by
\begin{widetext}
    \begin{gather}
        \bm{E}_\mathrm{t}(t)
        =\frac{t_{1\mathrm{s}}t_{2\mathrm{s}} e^{-i k_0 l_1}}{1 - r_{1\mathrm{s}}r_{4\mathrm{s}}r_{3\mathrm{s}}r_{2\mathrm{s}} e^{-i k_0 2(l_1+l_2)}} E_0 e^{i \omega_0 t}
        \begin{pmatrix}
        \bm{e}_\mathrm{s} & \bm{e}_\mathrm{p} \\
        \end{pmatrix}
        \begin{pmatrix}
        1 \\
        - \delta \phi(t) \\
        \end{pmatrix}, \\
        \delta \phi(t)
        \equiv \int_{-\infty}^\infty \frac{d\omega}{2\pi} \Tilde{\delta c}(\omega) e^{i \omega t} H_a(\omega),
    \end{gather}
\end{widetext}
where $l_1$ is the long side of the cavity length, $l_2$ is the short side of the cavity length, $E_0$ is the electric field of the laser light, $\omega_0$ is the angular frequency of the laser light, $\bm{e}_\mathrm{s}$ and $\bm{e}_\mathrm{p}$ are the basis vectors for s- and p-polarizations, respectively, $\delta\phi(t)$ is the polarization rotation angle of the transmitted light, $\Tilde{\delta c}(\omega)$ is the Fourier transform of $\delta c(t) \equiv \delta c_0 \sin{\qty(m_a t + \delta_{\tau}(t))}$, and $H_a(\omega)$ is the transfer function from $\Tilde{\delta c}(\omega)$ to the Fourier transform of $\delta\phi(t)$.
When s-polarization is resonant in the cavity, $k_0 2(l_1+l_2)=2n\pi$, the transfer function $H_a(\omega)$ is given by
\begin{widetext}
    \begin{align}
        H_a(\omega)
        = & k_0 \sqrt{\frac{1 - \qty|r_{2\mathrm{p}}|^2}{1 - \qty|r_{2\mathrm{s}}|^2}} \frac{1}{i \omega \qty(1 - r_{1\mathrm{p}}r_{4\mathrm{p}}r_{3\mathrm{p}}r_{2\mathrm{p}} e^{-i \omega 2(l_1+l_2)})} \left[- \qty(1 - e^{-i \omega l_2}) \qty(r_{1\mathrm{p}}r_{4\mathrm{s}}r_{3\mathrm{s}}r_{2\mathrm{s}} + r_{1\mathrm{p}}r_{4\mathrm{p}}r_{3\mathrm{p}}r_{2\mathrm{s}} e^{-i \omega (l_1+l_2)}) \right. \notag\\
        &+ \left. \qty(1 - e^{-i \omega l_1}) \qty(r_{1\mathrm{p}}r_{4\mathrm{p}}r_{3\mathrm{s}}r_{2\mathrm{s}} e^{-i \omega l_2} + r_{1\mathrm{p}}r_{4\mathrm{p}}r_{3\mathrm{p}}r_{2\mathrm{p}} e^{-i \omega (l_1+2l_2)}) \right].
    \end{align}
\end{widetext}
s- and p-polarizations receive reflection phase $\phi_{j\mathrm{s/p}}$ upon reflection on the $j$th mirror.
We define the reflection phase difference between s- and p-polarizations $\Delta\phi_j=\phi_{j\mathrm{s}}-\phi_{j\mathrm{p}}$, and assign this phase difference to $r_{j\mathrm{p}}$.
Therefore, $r_{j\mathrm{s}}$ is real, whereas $r_{j\mathrm{p}}$ is complex, expressed as $r_{j\mathrm{p}}=\abs{r_{j\mathrm{p}}} e^{i\Delta\phi_j}$.
$\Delta\phi_j$ is given by
\begin{equation}
    \Delta \phi_j = \frac{2\pi}{\nu_\mathrm{FSR}} \Delta \nu_j,
\end{equation}
where $\nu_\mathrm{FSR}=c/2(l_1+l_2)$, with $c$ being the speed of light, is the free spectral range (FSR) of the cavity, and $\Delta\nu_j=\nu_{j\mathrm{s}}-\nu_{j\mathrm{p}}$ is the resonant frequency difference between s- and p-polarizations.

In order to detect the polarization rotation generated by the axion-photon interaction, as shown in Fig.~\ref{fig:DANCE}, we use a half-wave plate (HWP) to leak a small amount of the s-polarization into the p-polarization, which acts as a local oscillator (LO) at angular frequency $\omega_0$~\cite{PhysRevD.100.023548}.
We aim to search for axion DM by detecting the beat signal between the LO and the p-polarization generated by the axion-photon interaction, which oscillates at angular frequency $\omega_0 \pm m_a$.
The sensitivity to the axion-photon coupling is limited by quantum shot noise~\cite{PhysRevLett.121.161301}.
The signal-to-noise ratio (SNR) is given by~\cite{PhysRevX.4.021030}
\begin{widetext}
    \begin{gather}
    \mathrm{SNR} = \dfrac{P_\mathrm{t,s}4\theta_\mathrm{HWP} \delta c_0 \qty|H^\prime_a(m_a)|}{\sqrt{2} \delta P_\mathrm{shot}} \times
    \begin{cases}
        T_\mathrm{obs}^{1/2}~(T_\mathrm{obs} < \tau), \\
        \qty(T_\mathrm{obs}\tau)^{1/4}~(\tau < T_\mathrm{obs}),
    \end{cases} \\
    P_\mathrm{t,s} \equiv \frac{\qty|t_{1\mathrm{s}}t_{2\mathrm{s}}|^2}{\qty|1 - r_{1\mathrm{s}}r_{4\mathrm{s}}r_{3\mathrm{s}}r_{2\mathrm{s}}|^2} \qty|E_0|^2, \\
    \delta P_\mathrm{shot} \equiv 2\theta_\mathrm{HWP} \sqrt{2\hbar \omega_0 P_\mathrm{t,s}}, \\
    \qty|H^\prime_a(m_a)| \equiv \frac{1}{2} \sqrt{(\Re \qty[H_a(m_a) + H_a(-m_a)])^2 + (\Im \qty[H_a(m_a) - H_a(-m_a)])^2},
    \end{gather}
\end{widetext}
where $P_\mathrm{t,s}$ is the power of the s-polarized transmitted light, $\theta_\mathrm{HWP}$ is the rotation angle of the HWP, $H^\prime_a(m_a)$ is the transfer function considering the case where the reflection phase difference between s- and p-polarizations is generated upon reflection on the mirrors, $\delta P_\mathrm{shot}$ is the one-sided power spectral density of the shot noise, $T_\mathrm{obs}$ is the observation time of axion DM, $\tau$ is the coherent time of axion DM, and $\hbar$ is the reduced Planck constant.
When $T_\mathrm{obs} < \tau$, the phase factor $\delta_{\tau}(t)$ stays constant during $\tau$, and the SNR scales as $T_\mathrm{obs}^{1/2}$.
In contrast, when $\tau < T_\mathrm{obs}$, $\delta_{\tau}(t)$ stays constant within each $\tau$, but changes randomly between different $\tau$.
Therefore, the SNR scales as $\qty(T_\mathrm{obs}\tau)^{1/4}$.

\begin{figure}
    \centering
    \includegraphics[width=1.0\linewidth]{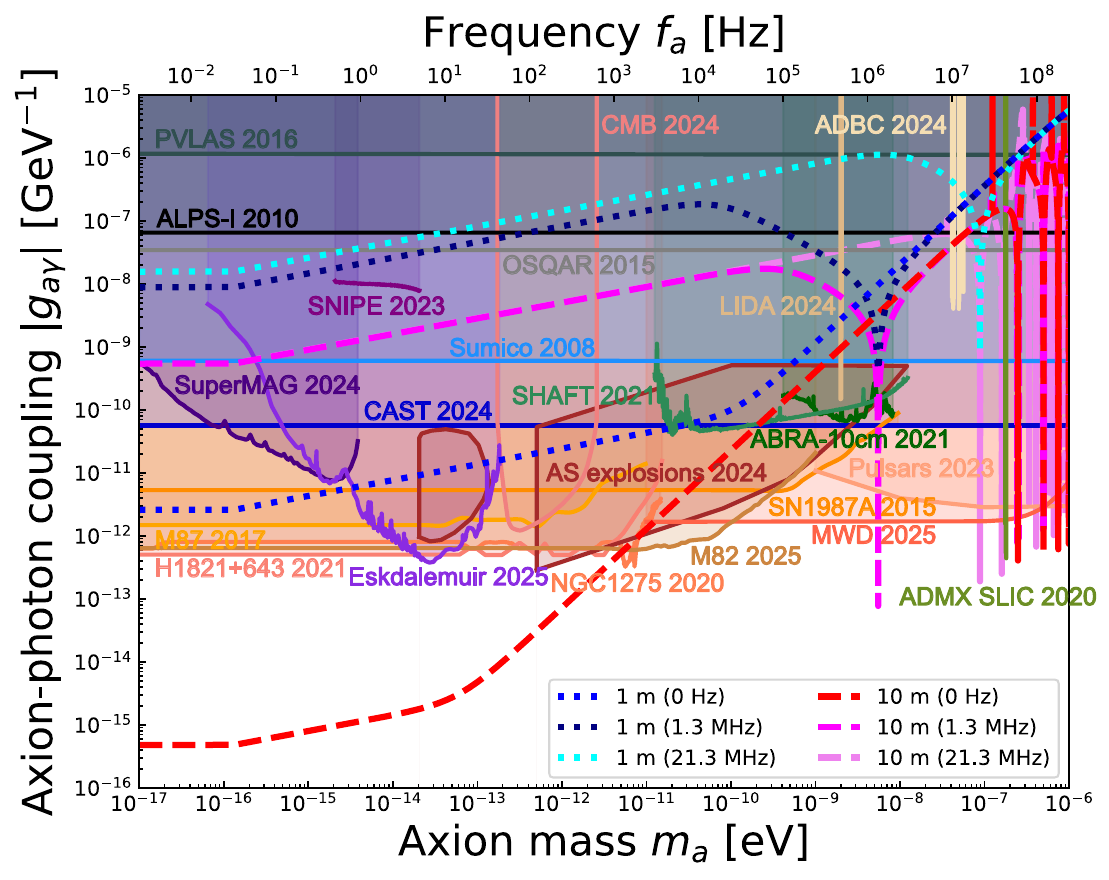}
    \caption{
    The target sensitivity to the axion-photon coupling of DANCE and the current upper limits.
    The dotted and dashed lines represent the target sensitivity based on the parameters listed in Tables~\ref{tab:mirrorparameter} and \ref{tab:DANCEparameter}.
    We assume that each mirror has the same reflection phase difference between s- and p-polarizations, $T_\mathrm{obs}=\SI{1}{year}$, and no optical loss.
    The blue, navy, and cyan dotted lines correspond to (a), (b), and (c) of Table~\ref{tab:mirrorparameter}, with the upper row of Table~\ref{tab:DANCEparameter}.
    The red, magenta, and violet dashed lines correspond to the same rows of Table~\ref{tab:mirrorparameter}, with the lower row of Table~\ref{tab:DANCEparameter}.
    The colored bands show the current upper limits obtained from ADMX SLIC~\cite{PhysRevLett.124.241101}, Sumico~\cite{INOUE_2008_93}, CAST~\cite{PhysRevLett.133.221005}, ALPS~\cite{EHRET_2010_149}, OSQAR~\cite{PhysRevD.92.092002}, PVLAS experiment~\cite{Valle_2016}, ABRACADABRA~\cite{PhysRevLett.127.081801}, SHAFT~\cite{Gramolin_2021}, SN1987A~\cite{Payez_2015}, M87~\cite{Marsh_2017}, NGC1275~\cite{Reynolds_2020}, H1821+643~\cite{H1821643_2021}, pulsars~\cite{PhysRevLett.131.111004}, axion star explosions~\cite{PhysRevD.109.043018}, CMB~\cite{PhysRevLett.134.081001}, MWD polarization~\cite{benabou2025searchaxionsmagneticwhite}, M82~\cite{PhysRevLett.134.171003}, SNIPE~\cite{PhysRevD.108.096026}, SuperMAG~\cite{PhysRevD.110.115036}, Eskdalemuir~\cite{nishizawa2025axiondarkmattersearch}, LIDA~\cite{PhysRevLett.132.191002}, and ADBC experiment~\cite{PhysRevLett.133.111003}.
    }
    \label{fig:sensitivity}
\end{figure}

Assuming $\mathrm{SNR} \ge 1$, the axion-photon coupling $\gag$ is given by
\begin{widetext}
    \begin{equation}
        \label{sensitivity}
        \gag \geq 1.55 \times 10^{-11}~\si{GeV^{-1}} \dfrac{\qty|1 - r_{1\mathrm{s}}r_{4\mathrm{s}}r_{3\mathrm{s}}r_{2\mathrm{s}}|}{\qty|t_{1\mathrm{s}}t_{2\mathrm{s}}|} \sqrt{\dfrac{\SI{0.4}{GeV~cm^{-3}}}{\rho_a} \dfrac{\SI{1}{W}}{P_\mathrm{in}} \dfrac{\SI{1064}{nm}}{\lambda_0}} \dfrac{4.0 \times 10^9~\si{eV}^{-1}}{\qty|H^\prime_a(m_a)|/{k_0}} \times
        \begin{cases}
        \qty(\dfrac{\SI{1}{year}}{T_\mathrm{obs}})^{1/2}~\qty(T_\mathrm{obs} < \tau), \\
        \qty(\dfrac{\SI{1}{year}}{T_\mathrm{obs} \tau})^{1/4}~\qty(\tau < T_\mathrm{obs}),
        \end{cases}
    \end{equation}
\end{widetext}
where $P_\mathrm{in}\equiv\qty|E_0|^2$ is the input laser power to the cavity and $\lambda_0$ is the laser wavelength.
Figure~\ref{fig:sensitivity} shows the target sensitivity to the axion-photon coupling of DANCE calculated using Eq.~\eqref{sensitivity} and the current upper limits obtained from previous research.
The target sensitivity is based on the parameters listed in Tables~\ref{tab:mirrorparameter} and \ref{tab:DANCEparameter}.
We assume that the reflection phase difference between s- and p-polarizations is identical for each mirror, $T_\mathrm{obs}=\SI{1}{year}$, and no optical loss, i.e., $\abs{r_{j\mathrm{s/p}}}^2+\abs{t_{j\mathrm{s/p}}}^2=1$.
When $\Delta\phi_j=0$, we set $\qty|H^\prime_a(m_a)|=\qty|H_a(m_a)|$, enabling a sensitive axion DM search in the low axion mass region.
Tuning the laser wavelength to \SI{1065}{nm} to satisfy $\Delta\phi_j=0$, simultaneous resonance can be achieved as shown by the blue dotted and red dashed lines in Fig.~\ref{fig:sensitivity}.
The first bending point in Fig.~\ref{fig:sensitivity} is determined by $T_\mathrm{obs}$ and $\tau$.
When $T_\mathrm{obs}<\tau$, $\gag$ is flat.
Conversely, when $\tau<T_\mathrm{obs}$, $\gag$ scales as $m_a^{1/4}$.
The second bending point in Fig.~\ref{fig:sensitivity} corresponds to the HWHM of the cavity for p-polarization, given by
\begin{equation}
    \label{requirement}
    \nu_\mathrm{HWHM,p} = \frac{c}{4(l_1+l_2) \mathcal{F}_\mathrm{p}},
\end{equation}
where $\mathcal{F}_{\mathrm{p}}$ is the finesse of the p-polarization
\begin{equation}
    \mathcal{F}_{\mathrm{p}}=\frac{\pi \sqrt{r_\mathrm{1p}r_\mathrm{4p}r_\mathrm{3p}r_\mathrm{2p}}}{1 - r_\mathrm{1p}r_\mathrm{4p}r_\mathrm{3p}r_\mathrm{2p}}.
\end{equation}
When the axion mass is higher than the mass corresponding to the $\nu_\mathrm{HWHM,p}$, $\gag$ scales as $m_a^{5/4}$.

On the other hand, when $\Delta\phi_j \neq 0$, we must use the modified transfer function $H^\prime_a(m_a)$, which degrades the sensitivity in the low axion mass region.
However, it is possible to enhance the sensitivity at the axion mass corresponding to the total resonant frequency difference between s- and p-polarizations $\Delta\nu_\mathrm{tot}$ by tuning the laser wavelength appropriately as shown in Fig.~\ref{fig:sensitivity}.
The reflectivity of mirrors depends on the laser wavelength.
Therefore, the wavelength needs to be tuned within the range where the reflectivity remains high to prevent the sensitivity degradation.
The wavelength can be adjusted up to \SI{1045}{nm}, which corresponds to the total reflection phase difference between s- and p-polarizations $\Delta\phi_\mathrm{tot}$ of \SI{-25.6}{deg}, as indicated by the cyan dotted and violet dashed lines in Fig.~\ref{fig:sensitivity}.

\begin{table}
    \centering
    \caption{
    Values of the laser wavelength $\lambda_0$, the total reflection phase difference between s- and p-polarizations $\Delta\phi_\mathrm{tot}$, and the total resonant frequency difference between two polarizations $\Delta\nu_\mathrm{tot}$ are shown.
    The values of $\Delta\phi_\mathrm{tot}$ and $\Delta\nu_\mathrm{tot}$ are based on the design specifications of the test mirror used in this work.
    }
    \begin{tabular}{cccc}
    \hline
    & $\lambda_0$ & $\Delta\phi_\mathrm{tot}$ & $\Delta\nu_\mathrm{tot}$ \\
    \hline
    (a) & \SI{1065}{nm} & \SI{0}{deg} & \SI{0}{Hz} \\
    (b) & \SI{1064}{nm} & \SI{-1.6}{deg} & \SI{-1.3}{MHz} \\
    (c) & \SI{1045}{nm} & \SI{-25.6}{deg} & \SI{-21.3}{MHz} \\
    \hline
    \end{tabular}
    \label{tab:mirrorparameter}
\end{table}

\begin{table}
    \centering
    \caption{
    Parameters used to calculate the target sensitivity of DANCE.
    The upper and lower rows correspond to the configurations with round-trip lengths of \SI{1}{m} and \SI{10}{m}, respectively.
    }
    \begin{tabular}{lcc}
    \hline
    Parameter & Symbol & Value \\
    \hline
    Round-trip length & $2(l_1+l_2)$ & \SI{1}{m} \\
    ~~~Cavity length & $(l_1, l_2)$ & (\SI{45}{cm}, \SI{5}{cm}) \\
    Input power & $P_\mathrm{in}$ & \SI{1}{W} \\
    Transmissivity & & \\
    ~~~1st mirror & $\qty(\abs{t_{1\mathrm{s}}}^2, \abs{t_{1\mathrm{p}}}^2)$ & ($\SI{6}{ppm}$, $\SI{300}{ppm}$) \\
    ~~~2nd mirror & $\qty(\abs{t_{2\mathrm{s}}}^2, \abs{t_{2\mathrm{p}}}^2)$ & ($\SI{6}{ppm}$, $\SI{300}{ppm}$) \\
    ~~~3rd mirror & $\qty(\abs{t_{3\mathrm{s}}}^2, \abs{t_{3\mathrm{p}}}^2)$ & ($\SI{6}{ppm}$, $\SI{300}{ppm}$) \\
    ~~~4th mirror & $\qty(\abs{t_{4\mathrm{s}}}^2, \abs{t_{4\mathrm{p}}}^2)$ & ($\SI{6}{ppm}$, $\SI{300}{ppm}$) \\
    \hline
    Round-trip length & $2(l_1+l_2)$ & \SI{10}{m} \\
    ~~~Cavity length & $(l_1, l_2)$ & (\SI{4.5}{m}, \SI{0.5}{m}) \\
    Input power & $P_\mathrm{in}$ & \SI{100}{W} \\
    Transmissivity & & \\
    ~~~1st mirror & $\qty(\abs{t_{1\mathrm{s}}}^2, \abs{t_{1\mathrm{p}}}^2)$ & ($\SI{3.14}{ppm}$, $\SI{3.14}{ppm}$) \\
    ~~~2nd mirror & $\qty(\abs{t_{2\mathrm{s}}}^2, \abs{t_{2\mathrm{p}}}^2)$ & ($\SI{3.14}{ppm}$, $\SI{3.14}{ppm}$) \\
    Reflectivity & & \\
    ~~~3rd mirror & $\qty(\abs{r_{3\mathrm{s}}}^2, \abs{r_{3\mathrm{p}}}^2)$ & (100\%, 100\%) \\
    ~~~4th mirror & $\qty(\abs{r_{4\mathrm{s}}}^2, \abs{r_{4\mathrm{p}}}^2)$ & (100\%, 100\%) \\
    \hline
    \end{tabular}
    \label{tab:DANCEparameter}
\end{table}

\section{Experiment}
\label{sec:experiment}
\begin{figure}
    \centering
    \includegraphics[width=1.0\linewidth]{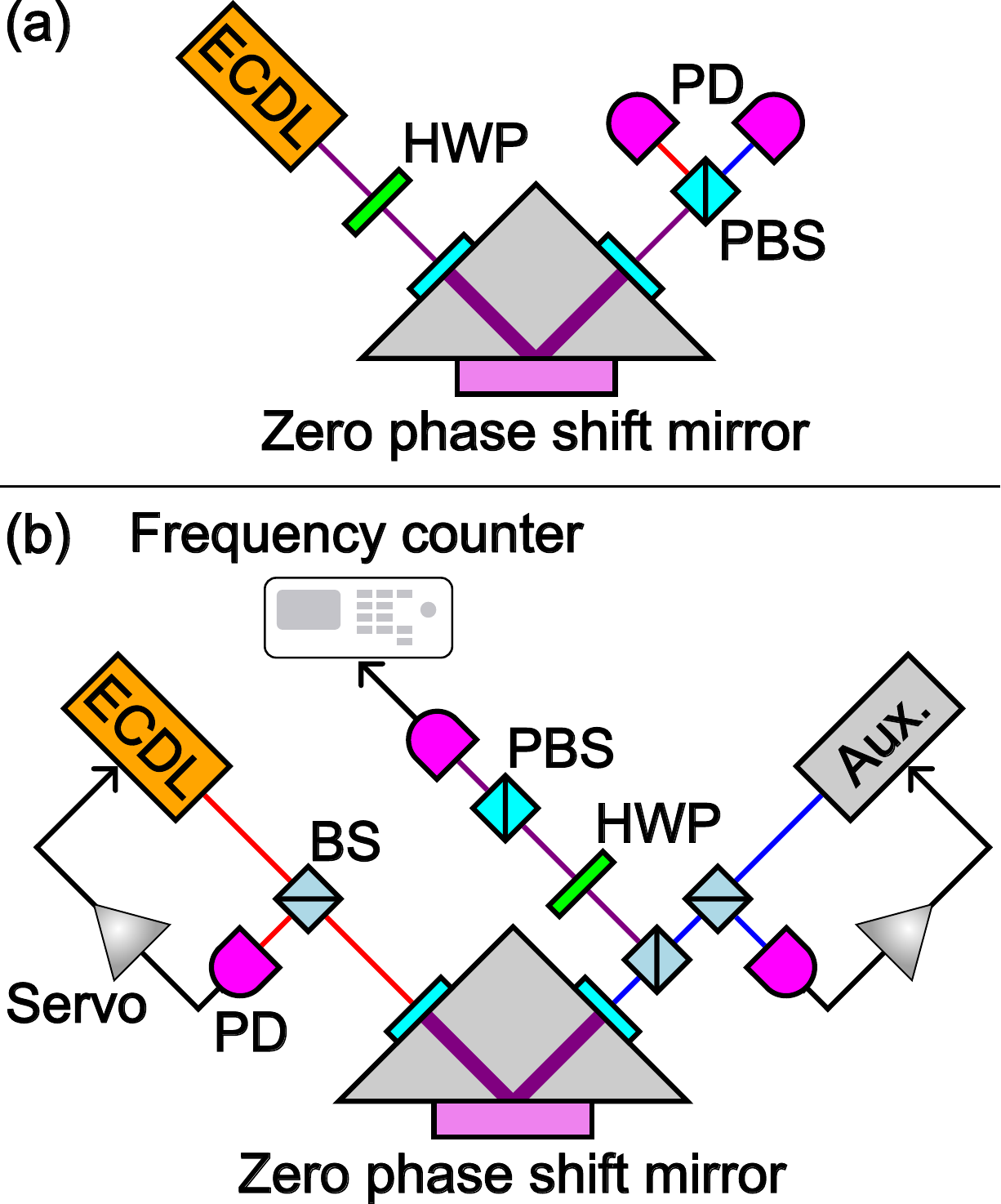}
    \caption{
    Schematic of the experimental setup.
    The red and blue lines represent s- and p-polarizations, respectively.
    The purple line indicates a mixture of s- and p-polarizations.
    (a) Proof of principle for simultaneous resonance.
    (b) Measurement of the fluctuations of the beat note between s- and p-polarizations.
    ECDL: external cavity diode laser,
    BS: beam splitter.
    }
    \label{fig:setup}
\end{figure}

We describe the concept and setup of our experiment.
According to~\cite{PhysRevD.108.072005}, the reflection phase difference between s- and p-polarizations of each mirror was estimated by dividing the total phase difference from four mirrors by four.
In this method, we did not accurately estimate the phase difference.
In order to address this issue, we proposed a folded cavity to measure the phase difference as shown in Fig.~\ref{fig:setup}.
The cavity is made of a super-invar spacer and consists of the input, end, and test mirrors, all of which are fixed to the spacer by jigs.
Table~\ref{tab:mirror} shows the summary of the folded cavity parameters.
Both the input and end mirrors are direct incidence.
Incident angle on the test mirror is \SI{42}{deg}, which is the same as that used in~\cite{PhysRevD.108.072005}.
Therefore, we can measure the reflection phase difference between s- and p-polarizations generated only on the test mirror.
Using this setup, we can accurately estimate the phase difference for each mirror in a bow-tie ring cavity, thereby improving the calibration accuracy of the sensitivity to the axion-photon coupling.

Long-term measurements of DANCE improve the sensitivity.
However, if the reflection phase difference between s- and p-polarizations fluctuates during the measurement, the sensitivity will degrade.
The requirement for the phase difference per mirror $\Delta \phi_\mathrm{req}$, imposed by DANCE with a round-trip length of \SI{1}{m} is given by
\begin{equation}
    \abs{\Delta \phi_\mathrm{req} - \Delta \phi_\mathrm{ave}} \le \frac{1}{4}\frac{\nu_\mathrm{HWHM,p}}{\nu_\mathrm{FSR}} \times 360~\si{deg} = 8.6 \times 10^{-3}~\si{deg}.
\end{equation}
This requirement is determined under the assumption of no optical loss, using the transmissivity of p-polarization in the upper row of Table~\ref{tab:DANCEparameter} and applying Eq.~\eqref{requirement}.
$\Delta\phi_\mathrm{ave}$ is the average value of the reflection phase difference between s- and p-polarizations.
Satisfying this requirement suggests that simultaneous resonance is achievable.

In this work, we used a zero phase shift mirror with a wavelength-dependent reflection phase difference between s- and p-polarizations of \SI{0.34}{deg/nm}.
The phase difference of the mirror is \SI{0}{deg} at \SI{1065}{nm}.
However, due to manufacturing errors, achieving this exact value is generally difficult.
Therefore, it is necessary to use a wavelength tunable laser which can vary the laser wavelength over a wide range, as the cavity is expected to achieve simultaneous resonance at a wavelength different from the design specification.
Thus, we employed an external cavity diode laser (ECDL)~\cite{Zorabedian:88, BAILLARD_2006_609}.
Its wavelength tuning range is $\qtyrange{1038}{1068}{nm}$.

We conducted two types of experiments to investigate the issues mentioned above.
The first experiment aims to probe a wavelength which is able to achieve simultaneous resonance.
As shown in the upper figure in Fig.~\ref{fig:setup}, the incident laser beam into the cavity was adjusted using a HWP to achieve a 50:50 ratio of s- and p-polarizations.
The injected laser power was \SI{10}{mW}.
Applying a triangular wave to the piezo-electric transducer (PZT) of the laser source, we used a PBS to separate the transmitted light into s- and p-polarizations, which were detected with two PDs.
The resonant frequency difference between s- and p-polarizations was estimated by fitting the resonant peak with a Lorentzian function.
The calibration from the frequency difference $\Delta\nu$ to the reflection phase difference between s- and p-polarizations $\Delta\phi$ is given by
\begin{equation}
    \label{calibration}
    \Delta \phi = \frac{\Delta \nu}{\nu_{\mathrm{FSR,folded}}} \times \SI{180}{deg},
\end{equation}
where $\nu_{\mathrm{FSR,folded}}=c/2l$ is the FSR of the folded cavity, $l$ is the cavity length of the folded cavity.
Since the laser beam reflects twice on the test mirror during a round-trip inside the cavity, it is necessary to divide \SI{360}{deg} by two.
We conducted the above procedure at ten wavelengths in the range of approximately $\qtyrange{1064}{1068}{nm}$.

The second experiment aims to evaluate the fluctuations of the reflection phase difference between s- and p-polarizations through 24-hour measurement.
The lower figure in Fig.~\ref{fig:setup} shows the experimental setup.
In addition to the ECDL, we employed an auxiliary laser to obtain time series data of the beat note between s- and p-polarizations.
The polarizations of the ECDL and the auxiliary laser were adjusted to s- and p-polarizations, respectively.
The power of the laser beams injected into the cavity was \SI{8}{mW} for s-polarization and \SI{4}{mW} for p-polarization.
After injecting both laser beams into the cavity, we tuned the wavelengths of both lasers to be identical.
However, the wavelength tuning range of the auxiliary laser is limited to \SI{0.3}{nm} around \SI{1064}{nm}, so we set the wavelength to \SI{1064.27}{nm}.
Each laser frequency was locked to $\mathrm{TEM}_{00}$ mode using the Pound-Drever-Hall method~\cite{Drever_1983}.
In order to suppress laser frequency noise, we designed high-gain filters.
The estimated open-loop transfer function gains below \SI{1}{Hz} are $G_{\mathrm{s}} \sim 7.5 \times 10^{5}$ for s-polarization and $G_{\mathrm{p}} \sim 5.3 \times 10^{5}$ for p-polarization.
The transmitted light of s-polarization from the ECDL and the reflected light of p-polarization from the auxiliary laser passed through a HWP, and interfered with each other after being projected onto the p-polarization axis with a PBS.
In this setup, signals from other frequency bands, such as the modulation frequency \SI{15}{MHz} used for frequency control, were mixed into the beat note, distorting its time series data.
In order to address this, we used a band-pass filter consisting of passive high-pass and low-pass filters to eliminate unwanted signals, and measured the beat note between s- and p-polarizations with a frequency counter.
We recorded the beat note at a sampling rate of \SI{10}{Hz} for 24 hours.

\begin{table}
    \centering
    \caption{
    Summary of the folded cavity parameters, all of which are based on the design specifications.
    }
    \begin{tabular}{lcc}
    \hline
    Parameter & Symbol & Value \\
    \hline
    Cavity length & $l$ & \SI{6}{cm} \\
    Reflectivity & & \\
    ~~~Input mirror & $\qty(\abs{r_\mathrm{input,s}}^2, \abs{r_\mathrm{input,p}}^2)$ & (99\%, 99\%) \\
    ~~~End mirror & $\qty(\abs{r_\mathrm{end,s}}^2, \abs{r_\mathrm{end,p}}^2)$ & (99\%, 99\%) \\
    Transmissivity & & \\
    ~~~Test mirror & $\qty(\abs{t_\mathrm{test,s}}^2, \abs{t_\mathrm{test,p}}^2)$ & (\SI{6}{ppm}, \SI{300}{ppm}) \\
    Radius of curvature & & \\
    ~~~Input mirror & $R_\mathrm{input}$ & \SI{50}{mm} \\
    ~~~End mirror & $R_\mathrm{end}$ & \SI{50}{mm} \\
    ~~~Test mirror & $R_\mathrm{test}$ & \SI{1000}{mm} \\
    Finesse & & \\
    ~~~s-polarization & $\mathcal{F}_{\mathrm{s,folded}}$ & 312.4 \\
    ~~~p-polarization & $\mathcal{F}_{\mathrm{p,folded}}$ & 303.5 \\
    \hline
    \end{tabular}
    \label{tab:mirror}
\end{table}

\section{Results}
\label{sec:results}
\begin{figure}
    \centering
    \includegraphics[width=1.0\linewidth]{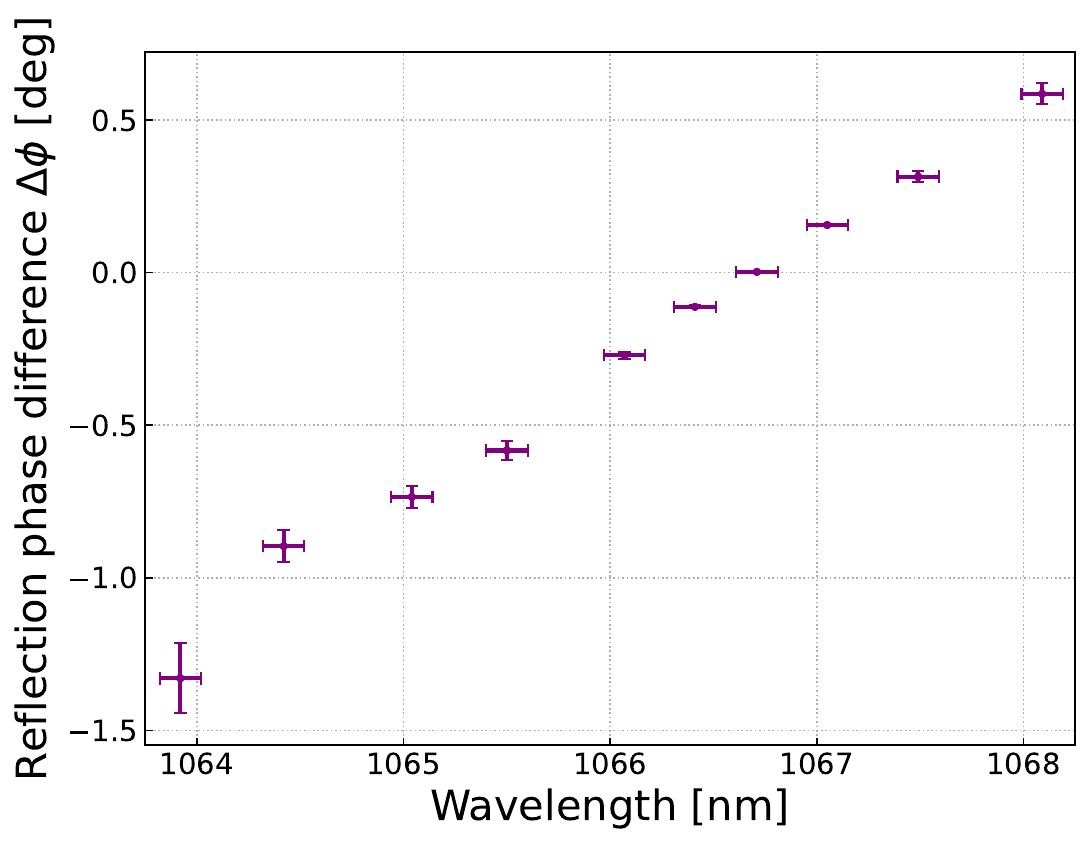}
    \caption{
    Wavelength dependence of the reflection phase difference between s- and p-polarizations.
    }
    \label{fig:phasedifference}
\end{figure}

We present the results of the measurement of the reflection phase difference between s- and p-polarizations.
Figure~\ref{fig:phasedifference} shows the results of wavelength dependence of the phase difference.
The error bar on the wavelength represents the resolution of the spectrometer used to measure the laser wavelength, and that on the phase difference indicates the statistical error estimated from ten measurements.
The phase difference was $(2.3 \pm 1.1) \times 10^{-3}~\si{deg}$ at \SI{1066.7}{nm} and satisfied the requirement of $8.6 \times 10^{-3}~\si{deg}$.
The wavelength was found to be \SI{1.7}{nm} higher than the design specification.
The wavelength dependence of the reflection phase difference between s- and p-polarizations is nearly consistent with the design specification.

\begin{figure}
    \centering
    \includegraphics[width=1.0\linewidth]{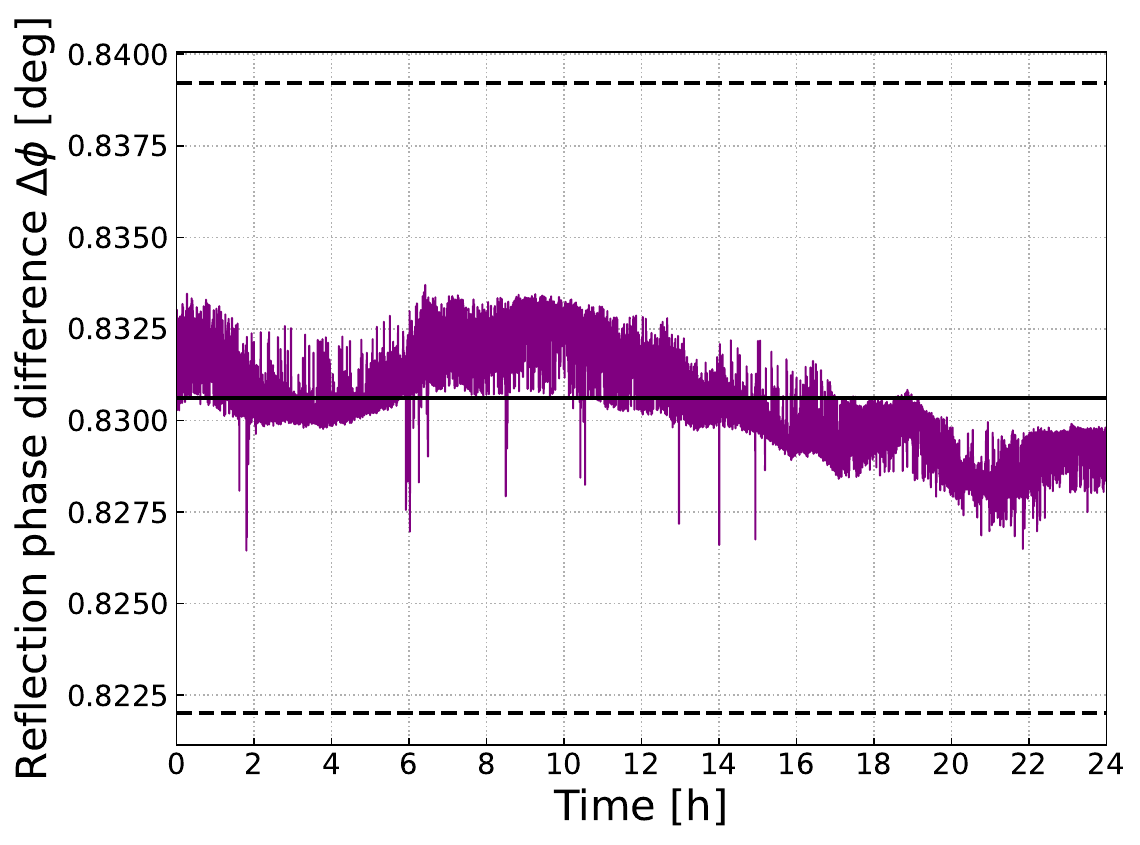}
    \caption{
    Time series data of the beat note between s- and p-polarizations.
    The black solid line indicates the average reflection phase difference between s- and p-polarizations $\Delta\phi_\mathrm{ave}=0.8306~\si{deg}$, and the black dashed line represents the requirement range for the phase difference, $\Delta\phi_\mathrm{ave} \pm 8.6 \times 10^{-3}~\si{deg}$.
    }
    \label{fig:beat}
\end{figure}

Figure~\ref{fig:beat} shows the time series data of the beat note between s- and p-polarizations, calculated using Eq.~\eqref{calibration}.
The average reflection phase difference between s- and p-polarizations was $\Delta\phi_\mathrm{ave}=0.8306~\si{deg}$, and the standard deviation was $1.3 \times 10^{-3}~\si{deg}$.
This fluctuation is 6.6 times smaller than the requirement of $8.6 \times 10^{-3}~\si{deg}$.

\section{Discussion}
\label{sec:discussion}
We discuss the results of this work.
The laser wavelength which satisfies the requirement for the reflection phase difference between s- and p-polarizations was shifted from the design specification as shown in Fig.~\ref{fig:phasedifference}.
We consider this shift to be attributed to the thickness error in the dielectric multilayer coating of the zero phase shift mirror.

Figure~\ref{fig:beat} suggests that simultaneous resonance could be stable for 24 hours.
Note that achieving simultaneous resonance requires that all four mirrors used in the bow-tie ring cavity satisfy the requirement for the reflection phase difference between s- and p-polarizations per mirror.
Here, we focus on discussing the cause of the beat note fluctuations.
Although common mode rejection was expected to be effective, the measured fluctuations were approximately three orders of magnitude larger than the frequency noise.
In principle, the fluctuations can be limited by the frequency counter noise.
However, they were approximately six orders of magnitude worse than the estimated frequency counter noise.
We consider room temperature fluctuations to be one of the causes.
They could cause thickness fluctuations in the dielectric multilayer coating of the zero phase shift mirror, inducing the fluctuations of the reflection phase difference between s- and p-polarizations.

In addition to thickness fluctuations, nonuniformities and spatial inhomogeneities in the mirror coating could lead to fluctuations of the phase difference across the beam spot.
If these effects occur in multiple mirrors, they would change the wavelength dependence of the phase difference and increase the fluctuations of the phase difference.
Moreover, large manufacturing tolerances in the mirror coating could cause similar problems.
Due to the above factors, the reflection phase difference between s- and p-polarizations might not satisfy the requirement, possibly preventing simultaneous resonance.

In this work, the sampling rate for measuring the beat note was \SI{10}{Hz}, and thus beat notes at frequencies above \SI{10}{Hz}, which are also within the range targeted by DANCE, were not explored.
A possible noise source in this frequency band is fluctuations of the resonant frequency difference between s- and p-polarizations.
If the incident polarization into the bow-tie ring cavity is elliptical, fluctuations in the cavity length can couple to the incident laser beam, potentially causing the frequency difference to fluctuate.
Therefore, the sensitivity to the axion-photon coupling could be limited by this noise.

\section{Conclusion}
\label{sec:conclusion}
Novel experiments using optical cavities have been proposed.
The first demonstration with the ring cavity has already been conducted, revealing that the method limits the sensitivity to the axion-photon coupling in the low axion mass region due to the reflection phase difference between s- and p-polarizations upon reflection on the mirror~\cite{PhysRevLett.132.191002, PhysRevLett.133.111003, PhysRevD.108.072005}.
In this paper, we propose a new approach for a broadband and sensitive axion DM search.
In the low axion mass region, we can improve the sensitivity by achieving simultaneous resonance with a zero phase shift mirror and a wavelength tunable laser.
In the high axion mass region, it is possible to search under nonsimultaneous resonance by tuning the laser wavelength such that the reflection phase difference between s- and p-polarizations corresponds to the axion mass at which a peak appears.

In this work, we confirmed that the reflection phase difference between s- and p-polarizations satisfied the requirement imposed by DANCE with a round-trip length of \SI{1}{m}.
Therefore, we demonstrated a proof of principle for simultaneous resonance and successfully evaluated the phase difference per mirror using the folded cavity, the zero phase shift mirror, and the wavelength tunable laser.
The mirror evaluation method developed in this work improves the calibration accuracy for the sensitivity in axion DM searches using ring cavities.

\section*{Acknowledgment}
We would like to thank Shigemi Otsuka and Togo Shimozawa for manufacturing some parts, Layertec GmbH for fabricating the custom-designed mirrors, and Kenichi Nakagawa for providing detailed instructions on the use and assembly of the ECDL.
This work is supported by JSPS KAKENHI Grant Nos. 19K14702, 20H05850, 20H05854, 20H05859, 20H05639, 24K00640, and 24K21546, and by JST FOREST Program No. JPMJFR222Y.

\section*{Data availability}
The data are not publicly available.
The data are available from the authors upon reasonable request.

\bibliography{reference}
\bibliographystyle{apsrev4-2}

\end{document}